\documentclass[onecolumn,aps,prd,superscriptaddress,preprintnumbers,nofootinbib,10pt]{revtex4-2}
\usepackage{amsfonts}
\usepackage[tbtags]{amsmath}
\usepackage{amssymb}
\usepackage{color}
\usepackage{courier}
\usepackage{dsfont}
\usepackage{euscript}
\usepackage{epsfig}
\usepackage{epstopdf}
\usepackage{float}
\usepackage{graphics}
\usepackage{subcaption}
\usepackage{graphicx}
\usepackage[utf8]{inputenc}
\usepackage{latexsym}
\usepackage{MnSymbol}
\usepackage{pifont}
\usepackage{physics}
\usepackage{slashed}
\usepackage[svgnames]{xcolor}
\usepackage[normalem]{ulem}
\usepackage{url}
\usepackage{verbatim}
\usepackage{widetable}
\usepackage{setspace}
\usepackage[plainpages=false,pdfpagelabels]{hyperref}
\usepackage{orcidlink}

\allowdisplaybreaks

\hypersetup{
	colorlinks=true,
	citecolor=blue,
	citebordercolor=red,
	linktoc=all,
	linkcolor=blue,
	urlcolor=blue
}

\newcommand{\ee}[1]{e^{#1}}

\begin{document}

	\title{\bf \Large More on Majorana fields, modular localization and near saturation of the Tsirelson bound}
	
		\vspace{1cm}

	\author{J. R. Apfata}\email{jeffersonr.apfata@gmail.com} \affiliation{UERJ -- Universidade do Estado do Rio de Janeiro,	Instituto de Física -- Departamento de Física Teórica -- Rua São Francisco Xavier 524, 20550-013, Maracanã, Rio de Janeiro, Brazil}

	\author{J. G.A. Caribé~\orcidlink{0000-0001-7418-5149}}\email{joaogcaribe@gmail.com} \affiliation{UERJ -- Universidade do Estado do Rio de Janeiro,	Instituto de Física -- Departamento de Física Teórica -- Rua São Francisco Xavier 524, 20550-013, Maracanã, Rio de Janeiro, Brazil}

	\author{S. P. Sorella~\orcidlink{0000-0002-6051-6960}} \email{silvio.sorella@fis.uerj.br} \affiliation{UERJ -- Universidade do Estado do Rio de Janeiro,	Instituto de Física -- Departamento de Física Teórica -- Rua São Francisco Xavier 524, 20550-013, Maracanã, Rio de Janeiro, Brazil}

	\begin{abstract}

The free massless Majorana field in $1+1$ dimensions is employed to study the Bell-CHSH inequality in Quantum Field Theory. Use of the modular wedge  localization enables us to show the near saturation of the Tsirelson bound.

\end{abstract}
		
\maketitle	
		
\section{Introduction}\label{sect}

The study of the Bell-CHSH inequality \cite{Bell:1964kc,Clauser:1969ny} in relativistic Quantum Field Theory is receiving increasing interest, both from theoretical, see \cite{Guimaraes:2024mmp} for an overview, as well as from experimental \cite{BESIII:2025vsr}
 and phenomenological sides \cite{Barr:2024djo}. \\\\The seminal results obtained by Summers-Werner \cite{Summers:1987fn,Summers:1987squ} have shown that the Bell-CHSH inequality is maximally violated in the vacuum state of free fields localized in wedge regions. This statement highlights the high degree of entanglement exhibited by the correlation functions of a Quantum Field Theory, with applications ranging from Quantum Gravity to information theory \cite{Witten:2018zxz}. \\\\Aiming to give an explicit example of how maximal violation is achieved in Quantum Field Theory, it seems safe to state that, till now, Fermi fields offer the simplest options, which can be worked out in fully analytic  form \cite{Dudal:2023mij,Dudal:2026eil,Caribe:2026mam,Caribe:2026xoa,Caribe:2026xoa}. This is due to the anti-commutation relations which, unlike the bosonic case, allow the introduction of genuine Hermitian, bounded, dichotomic operators built out from the spinor field itself. In this work we pursue the study of the role played by the Majorana fields in the violation of the Bell-CHSH inequality\footnote{See also the brief summary given in Appendix B of 
\cite{Caribe:2026mam}.}. We shall focus on the case of a free massless Majorana spinor. By relying on the introduction of the rapidity variable $\theta$, the 1-particle Hilbert space of the theory turns out to be given by $L^2(d \theta, {\mathbb{R}}) \oplus L^2(d \theta, {\mathbb{R}})$.\\\\Furthermore, by making use of the Bisognano-Wichmann results for wedge regions \cite{Bisognano:1975ih}, the modular theory of Tomita-Takesaki \cite{Takesaki:1970aki,Bratteli:1979tw,Summers:2003tf,Guido:2008jk,Brunetti:2002nt,Borchers:2000pv,Schroer:2014kya} can be introduced by following the setup already outlined  in the massive case \cite{Caribe:2026tfe}. This possibilitates the construction of analytic solutions for the maximal violation of the Bell-CHSH inequality by means of the modular wedge localization \cite{Brunetti:2002nt,Borchers:2000pv,Schroer:2014kya}. \\\\The paper is organized as follows. In Sect.\eqref{sect2} we review the canonical quantization of the massless Majorana field in $1+1$ Minkowski spacetime. In Sect.\eqref{sect3} we introduce the smearing procedure and identify the 1-particle space as $L^2(d \theta, {\mathbb{R}}) \oplus L^2(d \theta, {\mathbb{R}})$. Sect.\eqref{sect4} is devoted to the modular theory and to the Bell-CHSH inequality. In Sect.\eqref{sect5} we discuss the near saturation of the Tsirelson bound \cite{Cirelson:1980ry}, {\it i.e.} $2 \sqrt{2}$, corresponding to the maximal violation. Sect.\eqref{sect6} contains our conclusion.

\section{The massless Majorana field in $1+1$ dimensions and its quantization}\label{sect2}\label{sect2}

Let us start with the action of the model,
\begin{equation} 
S =  \int_{\mathcal{M}} d^2x \;  i{\bar \psi} \gamma^\mu \partial_\mu \psi, \label{Mact}
\end{equation}
where $\mathcal{M}$ stands for the Minkowski spacetime with metric $g_{\mu\nu}=diag(1,-1)$, and adopt the chiral representation
\begin{equation} 
\gamma^0=\begin{pmatrix}
0 & 1 \\
1 & 0
\end{pmatrix}\quad\text{and}\quad
\gamma^1=\begin{pmatrix}
0 & 1 \\
-1 & 0
\end{pmatrix}\; \label{conv}
\end{equation}
of the Dirac matrices.
Let the field $\psi$ be a two-component spinor that satisfies the Majorana condition  
\begin{equation} 
\psi^{C} = \psi,  \label{mj1}
\end{equation}
where $\psi^{\mathrm{C}} = \mathcal{C} {\bar \psi}^T $ is the Majorana conjugate of the spinor field $\psi$ and ${\bar \psi}^T = \psi^{\dagger}\gamma^0$ is the adjoint spinor. In this expression, $\mathcal{C}$ stands for the charge conjugation operator which obeys
\begin{equation} 
\mathcal{C} = \gamma^1 \;, \qquad \mathcal{C}  \gamma^\mu \mathcal{C} ^{-1}= -(\gamma^{\mu})^T \;, \qquad \mathcal{C} ^2=-\mathds{1} \;. \label{chconj}
\end{equation} 

The Majorana condition \eqref{mj1} implies that the spinor field $\psi$ must be of the form
\begin{equation} 
\psi(t,x) = \begin{pmatrix}
h(t,x)  \\
i \varphi(t,x)
\end{pmatrix} \;, \label{mj2}
\end{equation}
where $h$ and $\varphi$ are real fields. 

The equations of motion for $\psi$ that result from extremizing the action \eqref{Mact} can be compactly written as
\begin{equation}
    \gamma^{\mu}\partial_\mu\psi = 0.\label{eq:EOM}
\end{equation}
Plugging the spinor \eqref{mj2} into this equation results in
\begin{equation} 
(\partial_t + \partial_x)\varphi =0 \quad \text{and} \quad (\partial_t - \partial_x) h = 0 \label{mj3}
\end{equation}
and it follows that
\begin{equation} 
h = h(x_+)\quad\text{and}\quad \varphi = \varphi(x_-) \;, \label{chmj}
\end{equation}
where $x_{\pm}=t \pm x$ denote the light cone coordinates.

To obtain plane wave solutions to Eq.~\eqref{eq:EOM}, we define
\begin{equation}
    u(\omega,k)=\begin{pmatrix}
                    u_1(\omega,k)\\
                    u_2(\omega,k)
    \end{pmatrix}
\end{equation}
and substitute
\begin{equation}
    \psi(t,x) = \ee{-ik_{\mu}x^{\mu}}u(\omega,k),
\end{equation}
where $k^\mu = (\omega,k)$ and $x^\mu = (t,x)$, into this equation to obtain
\begin{equation}
    \gamma^{\mu}k_\mu\psi = 0.
\end{equation}
Solving it yields the dispersion relation $\omega = \abs{k}$ and the solutions
\begin{equation}
    u_1 = 0,\,u_2(k) \neq 0,\text{ for } k > 0\quad\text{and}\quad u_1(k) \neq 0,\,u_2 = 0,\text{ for } k < 0,
\end{equation}
where we now consider $u(\omega,k) = u(k)$ due to the dispersion relation. This solution can be compactly written as
\begin{equation}
    u(k) = \begin{pmatrix}
                    u_1(k)\\
                    0
    \end{pmatrix}\theta(-k)
    + 
    \begin{pmatrix}
                    0\\
                    u_2(k)
    \end{pmatrix}\theta(k)
\end{equation}

Thus, for the plane wave expansion, one has 
\begin{eqnarray} 
h(x_+) & = & \int_{-\infty}^0 \frac{dk}{2\pi} \frac{1}{\sqrt{2 |k|}} \left(a_k e^{-ikx_+} + a^\dagger_k e^{i kx_+}  \right) \;, \nonumber \\
\varphi(x_-) & = & \int_{0}^\infty \frac{dk}{2\pi} \frac{1}{\sqrt{2 |k|}} \left(b_k e^{-ikx_-} + b^\dagger_k e^{i kx_-}  \right) \;, \label{pwex}
\end{eqnarray}
where $a_k$, $a^\dagger_k$, $b_k$, and $b^\dagger_k$ are annihilation and creation operators obeying the anti-commutation relations 
\begin{eqnarray} 
\{ a_k, a^\dagger_q\} & = & 4\pi |k| \delta(k-q) \;, \qquad \{ a_k, a_q\} =0 \;, \qquad \{ a^\dagger_k, a^\dagger_q\} =0 \;, \nonumber \\
\{ b_k, b^\dagger_q\} & = & 4\pi |k| \delta(k-q) \;, \qquad \{ b_k, b_q\} =0 \;, \qquad \{ b^\dagger_k, b^\dagger_q\} =0 \;, \nonumber \\
\{ a_k, b^\dagger_q\} & = & 0\;, \qquad \{ a_k, b_q\} =0 \;, \label{carab}
\end{eqnarray}
from which it follows that 
\begin{equation} 
\{ h(t,x), h(t,y) \} = \delta(x-y) \;, \qquad \{ \varphi(t,x), \varphi(t,y) \} = \delta(x-y) \;, \qquad \{h(t,x), \varphi(t,y) \} = 0\;. \label{carhvphi}
\end{equation}
Hence, the massless Majorana field describes two modes, corresponding, respectively, to $(a_k, a^\dagger_k)$ and $(b_k, b^\dagger_k)$.

\section{Smearing. 1-particle Hilbert space }\label{sect3}

Quantum fields are operator-valued distributions \cite{Haag:1992hx}. As such, they need to be smeared in order to be properly defined as operators acting on the Hilbert space of the theory. In the present case, owing to the Majorana condition~\eqref{mj1}, the smearing procedure is accomplished by introducing a two-component real smooth spinor test function 
\begin{equation} 
f= \begin{pmatrix}
f_1(t,x)  \\
i f_2(t,x) 
\end{pmatrix}  \; \label{ftest}
\end{equation} 
and defining the smeared Majorana field as
\begin{equation}
    \psi(f) = \int_{\mathcal{M}}dt dx f^{\dagger}\psi(t,x) = \int_{\mathbb{R}} dx_+\; f_1(x_+) h(x_+) + \int_{\mathbb{R}} dx_- \; f_2(x_-) \varphi(x_-), \; \label{smf}
\end{equation}
where $\mathcal{M}$ stands for the Minkowski spacetime,
\begin{equation}
    f_1(x_{+}) = \frac{1}{2}\int_{\mathbb{R}}dx_{-}f_1(x_+,x_-)\quad\text{and}\quad f_2(x_{-}) = \frac{1}{2}\int_{\mathbb{R}}dx_{+}f_2(x_+,x_-).
\end{equation}
Making use of the plane wave expansion \eqref{pwex}, expression  \eqref{smf} can be rewritten as 
\begin{equation} 
\psi(f) = \psi(f)^\dagger = a_f + a_f^\dagger + b_f + b_f^\dagger \;, \label{hermsmf}
\end{equation}
where $a_f$ and $b_f$ stand for the smeared operators 
\begin{equation} 
a_f = \int_{-\infty}^0 \frac{dk}{2\pi} \frac{1}{\sqrt{2 |k|}}\; f_1^*(k) a_k \quad\text{and}\quad
b_f = \int_0^{\infty} \frac{dk}{2\pi} \frac{1}{\sqrt{2 |k|}}\; f_2^*(k) b_k \label{smab}
\end{equation}
with
\begin{equation} 
f_1(k) = \int_{\mathbb{R}} dx_+ \; e^{ikx_+} f_1(x_+)\quad\text{and}\quad f_2(k) = \int_{\mathbb{R}} dx_- \; e^{ikx_-} f_2(x_-).
\end{equation}
It is then easy to show that 
\begin{equation}
\{a_f, a^\dagger_g\} = \int_{-\infty}^0  \frac{dk}{2\pi} f_1^*(k) g_1(k)\quad\text{and}\quad\{b_f, b^\dagger_g\} = \int_{0}^{\infty}  \frac{dk}{2\pi} f_2^*(k) g_2(k).\label{qinn}
\end{equation}

We can now proceed with the identification of the 1-particle Hilbert space, whose inner product is obtained from the two-point Wightman correlation function $\langle 0| \psi(f) \psi(g) |0 \rangle$. From \eqref{qinn}, it follows that  
\begin{equation} 
\langle 0| \psi(f) \psi(g) |0\rangle =\langle f|g\rangle = \int_{-\infty}^0  \frac{dk}{2\pi} f_1^*(k) g_1(k) + \int_{0}^{\infty}  \frac{dk}{2\pi} f_2^*(k) g_2(k) \;, \label{qinn11}
\end{equation}
where use has been made of 
\begin{equation} 
a_f|0\rangle = b_f|0\rangle = 0 \;. \label{vcc}
\end{equation}
At this stage, it is helpful to introduce the rapidity variable, {\it i.e.} 
\begin{eqnarray} 
k & =&  - \mu e^{-\theta} \;, \qquad \theta \in (-\infty, \infty) \;, \qquad k\in (-\infty,0] \;, \nonumber \\
k & =&   \mu e^{\theta} \;, \qquad \theta \in (-\infty, \infty) \;, \qquad k\in [0,\infty)  \;, \label{rap}
\end{eqnarray}
where $\mu$ stands for an arbitrary scale, needed to restore the right dimensionality. Moreover, as we shall see later on, the scale $\mu$ will cancel out automatically when considering the Bell-CHSH correlation function. Thus, dropping the irrelevant factor $2\pi$, for the inner product $\langle f|g\rangle$ one gets 
\begin{equation} 
\langle f| g\rangle = \mu \int_{-\infty}^\infty   d\theta e^{-\theta} f_1^*(\theta) g_1(\theta) + \mu \int_{-\infty}^\infty   d\theta e^{\theta} f_2^*(\theta) g_2(\theta)  \;. \label{innf}
\end{equation}
We remind now that, under a boost  $\Lambda_s$ with parameter $s$
\begin{equation} 
(p^{0})^{'} = p^0 \cosh(s) - p \sinh(s) \;, \qquad p'=p \cosh(s) - p^0 \sinh(s) \;, \label{sboost}
\end{equation}
a two-component spinor $f$ transforms as 
\begin{equation} 
{\cal U}(\Lambda_s) f(\theta) {\cal U}^{-1}(\Lambda_s) = e^{-\frac{s}{2} \gamma^0 \gamma^1 } f(\theta-s) \;, \label{ubs}
\end{equation}
we have 
\begin{eqnarray} 
f_1(\theta) & \rightarrow & e^{\frac{s}{2}} f_1(\theta - s) \;, \nonumber \\
f_2(\theta) & \rightarrow & e^{-\frac{s}{2}} f_2(\theta - s) \;. \label{f1f2}
\end{eqnarray}
Therefore, introducing the quantities 
\begin{equation} 
{\hat f}_1(\theta) = e^{-\frac{\theta}{2}} f_1(\theta) \;, \qquad {\hat f}_2(\theta) = e^{\frac{\theta}{2}} f_2(\theta) \label{hatf1f2}
\end{equation}
one sees that $({\hat f}_1, {\hat f}_2)$ transform precisely as scalars, {\it i.e.} 
\begin{eqnarray} 
{\hat f}_1(\theta) & \rightarrow & e^{-\frac{\theta}{2}}e^{\frac{s}{2}} f_1(\theta - s) = {\hat f}_1(\theta-s) \;, \nonumber \\
{\hat f}_2(\theta) & \rightarrow & e^{\frac{\theta}{2}}e^{-\frac{s}{2}} f_2(\theta - s) = {\hat f}_2(\theta-s) \;. \label{fsc}
\end{eqnarray}
Finally, for the inner product $\langle f|g\rangle$ one obtains 
\begin{equation} 
\langle f|g\rangle = \mu \int_{-\infty}^\infty d\theta {\hat f}_1^*(\theta) {\hat g}_1(\theta) + \mu \int_{-\infty}^\infty d\theta {\hat f}_2^*(\theta) {\hat g}_2(\theta) \;, \label{fninn}
\end{equation}
showing that the 1-particle Hilbert space of the massless Majorana field is 
\begin{equation} 
{\cal H}_{1-part} = L^2(d \theta, {\mathbb{R}}) \oplus L^2(d \theta, {\mathbb{R}}) \;. \label{1pH}
\end{equation}

\section{Modular theory and the Bell-CHSH inequality }\label{sect4}

In order to address the issue of the violation of the Bell-CHSH inequality, we have to specify the spacetime regions to which we shall refer. As underlined in the introduction we shall consider the wedge regions 
\begin{equation} 
W_R =\{(x,t), \; x > |t|\}\quad\text{and}\quad W_L =\{(x,t), \; -x > |t|\} \;. \label{wrwl}
\end{equation}
These regions are the causal complement of each other: any point of $W_R$ is spacelike with respect to all points of $W_L$ and vice-versa. The reason for choosing these regions relies on the results obtained by Bisognano-Wichmann \cite{Bisognano:1975ih}, who have given a complete characterization of the modular operator $\delta$ as well as of the modular conjugation $j$ in such regions. In fact, it turns out that for wedge regions the modular operator $\delta$ is related to the self-adjoint generator $K$ of the boosts by the relation 
\begin{equation} 
\delta = e^{-2 \pi K} \;, \qquad K = - i \frac{\partial}{\partial \theta} \;. \label{dK}
\end{equation}
Moreover, for the modular conjugation $j$, one has 
\begin{equation} 
j = R_3(\pi) (\mathcal{C}PT) \;, \label{modj}
\end{equation}
where $(\mathcal{C}PT)$ stands for the CPT operator and $R_3(\pi)$ is a rotation of $\pi$ around the x-axis. The modular operator $\delta$ is self-adjoint and has a continuous spectrum with eigenvalues 
\begin{equation} 
\lambda^2(\omega) = e^ {-2\pi \omega} \in \mathbb{R}_+ \;, \label{spect}
\end{equation}
where $\omega \in \mathbb{R}$ are the eigenvalues of $K$, namely 
\begin{equation} 
K \psi_\omega = \omega \psi_\omega \;, \label{omeig}
\end{equation}
with generalized eigenstates $\psi_\omega$ given by plane waves in rapidity space 
\begin{equation} 
\psi_\omega = \frac{1}{\sqrt{2\pi}} e^{i \omega \theta} \;, \qquad \langle \psi_\omega| \psi_{\omega'} \rangle = \frac{1}{2 \pi} \int_{\infty}^\infty d \theta e^{-i \theta(\omega - \omega')} = \delta(\omega-\omega') \;. \label{eig}
\end{equation}
In particular, for the action of the operator $\delta^{1/2}$ on a scalar quantity $\psi(\theta)$, we have 
\begin{equation}
\delta^{1/2} \psi(\theta) = \psi(\theta+i\pi) \;. \label{daction}
\end{equation}
On the other hand, the modular conjugation $j$ is anti-linear and acts on scalars as 
\begin{equation} 
j \psi(\theta) = \psi(\theta)^* \label{jacts}
\end{equation}
because the factor $R_3(\pi)$ from Eq.~\eqref{modj} is irrelevant in this case. 

Following \cite{Guido:2008jk}, out of the operators $\delta$ and $j$, one defines the unbounded anti-linear Tomita-Takesaki operator  $s$: 
\begin{equation} 
s = j \delta^{1/2} \;. \label{s-op}
\end{equation}
The operators $s$, $j$ and $\delta$ are equipped with the following properties \cite{Guido:2008jk}:
\begin{equation} 
s^2=1 \; \qquad j^2=1 \;, \qquad j \delta^{1/2} j = \delta^{-1/2} \;, \qquad s^\dagger s = \delta \;. \label{prop}
\end{equation}
From eqs.\eqref{daction},\eqref{jacts}, it follows that 
\begin{equation} 
s \psi(\theta) = (\psi(\theta+i \pi))^*.\label{s-act}
\end{equation}
Given this condition, in order to properly define the domain of both $s$ and $\delta$, we require that, besides belonging to $L_2(d\theta, \mathbb{R})$, the quantity $\psi(\theta)$ has to exhibit analytic continuation in the strip $\theta +iy, \; 0\le y\le\pi$. Within the framework of the Bell-CHSH inequality, the introduction of the Tomita-Takesaki operator relies on the beautiful and powerful idea of modular localization \cite{Guido:2008jk,Brunetti:2002nt,Borchers:2000pv,Schroer:2014kya}. One says that a vector $f(\theta)$ of the 1-particle Hilbert space is $W_R$-localized if 
\begin{equation} 
s f(\theta) = f(\theta) \;. \label{wrloc}
\end{equation}
Similarly, a vector $g(\theta)$ is $W_L$-localized if 
\begin{equation} 
s^{\dagger} g(\theta) = g(\theta) \;, \label{wlloc}
\end{equation}
with 
\begin{equation} 
s^\dagger = \delta^{1/2} j = j \delta^{-1/2} \;. 
\end{equation}
From Eq.\eqref{s-act}, it follows that the $W_R$-localization condition presented in Eq.\eqref{wrloc}, amounts to imposing
\begin{equation} 
\psi(\theta) = (\psi(\theta+i\pi))^* \;. \label{ancond}
\end{equation}

\subsection{The Bell-CHSH inequality}\label{inin}

We have now all ingredients to state the Bell-CHSH inequality in the vacuum state. The first goal is that of introducing Hermitian, bounded, dichotomic operators. As already underlined, this task can be accomplished by employing the spinor field itself \cite{Summers:1987fn,Summers:1987squ} presented in Eq.\eqref{smf}. From the anti-commutation relations, it follows that the operator 
\begin{equation} 
{\cal A}(f) = \frac{\psi(f)}{||f||} \;, \label{Af}
\end{equation}
fulfills all requirements: 
\begin{equation} 
{\cal A}(f)= {\cal A}(f)^\dagger\quad\text{and}\quad{\cal A}(f)^2=1 \;. \label{Afd}
\end{equation}

We remark that when $f$ is $W_R$-localized and $g$ is $W_L$-localized, the operators $\mathcal{A}(f)$ and $\mathcal{A}(g)$ anti-commute instead of commuting due to their fermionic nature. In order to obtain commuting observables we employ the fermion-parity twist as in~\cite{Caribe:2026tfe} and define the Klein-transformed observables
\begin{equation}
    \mathcal{B}(g) = i\Gamma\mathcal{A}(g),
\end{equation}
where $\Gamma\mathcal{A}(g)\Gamma = -\mathcal{A}(g)$. Here, $\Gamma = (-1)^N$ and $N$ is the fermion number operator. The operator $\mathcal{B}(g)$ is Hermitian, dichotomic and commutes with $\mathcal{A}(f)$.

Therefore, for the Bell-CHSH correlation function in the vacuum state, one writes 
\begin{equation} 
\langle 0| {\cal C} |0\rangle = \langle 0| \left( {\cal A}(f) + {\cal A}(f') \right){\cal B}(g)+  \left( {\cal A}(f) - {\cal A}(f') \right){\cal B}(g') |0\rangle  \;, \label{bbll}
\end{equation}
where $(f,f')$ and $(g,g')$ are, respectively, $W_R$ and $W_L$ localized, {\it i.e.} 
\begin{equation} 
sf = f \;, \qquad sf'=f'\;, \qquad s^\dagger g= g \;, \qquad s^\dagger g'= g' \;. \label{ffgg}
\end{equation}
Violations of the Bell-CHSH inequality occur whenever 
\begin{equation}
2 < |\langle 0| {\cal C}|0\rangle | \le 2 \sqrt{2} \;. \label{vls}
\end{equation}
From Eq.\eqref{qinn11}, it follows that the correlation function $\langle 0| {\cal C}|0\rangle $ can be expressed in terms of inner products of the 1-particle space, namely 
\begin{equation} 
\langle 0| {\cal C} |0\rangle = -i \left(      \frac{\langle f|g\rangle}{||f||.||g||}  +   \frac{\langle f'|g\rangle}{||f'||.||g||}    +   \frac{\langle f|g'\rangle}{||f||.||g'||}  -   \frac{\langle f'|g'\rangle}{||f'||.||g'||}  \right) \;, \label{bf1}
\end{equation}
where the $-i$ factor arises from
\begin{equation} 
\bra{0}\mathcal{A}(f)\mathcal{B}(g)\ket{0} = - i\bra{0}\mathcal{A}(f)\mathcal{A}(g)\ket{0}, \label{ifact}
\end{equation} 
for spacelike $(f,g)$.

It remains now to choose suitable vectors $f$,$f'$,$g$ and $g'$. To accomplish this task, we follow the procedure outlined in \cite{Caribe:2026tfe} and introduce the quantity $\phi(\theta)$ defined through the half-sided Fourier transformation
\begin{equation} 
\phi(\theta) = \int_0^\infty d\omega \;h(\omega) e^{i \theta \omega} \;, \label{phith}
\end{equation}
where $h(\omega)$ is a real smooth function. It is straightforward to verify that expression \eqref{phith} exhibits an analytic continuation in the upper-half complex plane, thus fulfilling the condition needed for the well-definedness of the operators $s$ and $\delta$. Concerning now the choice of $(f,f')$, we set
\begin{equation}
    f_1 = f_2 = (1+s) \phi,\quad f'_1 = f'_2 = (1+s) i\phi,\quad g_1 = g_2 = \frac{\tilde{g} - \tilde{g}'}{\sqrt{2}},\quad\text{and}\quad g'_1 = g'_2 = \frac{\tilde{g} + \tilde{g}'}{\sqrt{2}}
\end{equation}
where\footnote{In the following, $\tilde{g}$ and $\tilde{g}'$ are not themselves $W_L$-localized. Instead, they are given by $i$ times a $W_L$-localized vector. This is done to take into account the twisted duality of the fermion fields~\cite{Summers:1987squ}.}
\begin{equation} 
\tilde{g} = i (1+s^\dagger) \phi\quad\text{and}\quad \tilde{g}'= -i (1+s^\dagger) i \phi \;. \label{hatggp}
\end{equation}
Using
\begin{equation}
    s \phi =  \int_0^\infty d\omega \;h(\omega) e^{-i \omega\theta} e^{-\pi \omega}\quad\text{and}\quad s^\dagger \phi =  \int_0^\infty d\omega \;h(\omega) e^{-i \omega\theta} e^{\pi \omega},
\end{equation}
one obtains a simple expression for the Bell-CHSH correlation function
\begin{equation} 
\langle 0| {\cal C} |0\rangle = 2\sqrt{2} \frac{2 \int_0^\infty d\omega \; h(\omega)^2  }{\sqrt{\int_0^\infty d\omega \; h(\omega)^2\left(1 + e^{-2\pi \omega} \right) }{\sqrt{\int_0^\infty d\omega \; h(\omega)^2\left(1 + e^{2\pi \omega} \right) }} } \; , \label{bellf1}
\end{equation} 
showing that the violation of the Bell-CHSH inequality is encoded in a judicious choice of the function $h(\omega)$.

\section{Semi-analytic solution with near saturation of the Tsirelson bound}\label{sect5}

As we have seen in the previous section, the framework of the modular theory has enabled us to express the whole Bell-CHSH correlation function in terms of a unique quantity: $h(\omega)$. This feature facilitates the search for simple analytic solutions. As a concrete example, we might rely on a two parameter expression given by 
\begin{equation} 
h(\omega)^2 = \frac{a^2}{\omega^2+a^2} \;e^{-\frac{\omega^2}{b}} \;, \qquad b>0 \;, \label{abt}
\end{equation}
where $(a,b)$ stand for arbitrary real parameters to be chosen by requiring maximal violations. The rationale behind expression \eqref{abt} follows from an important observation made by Summers-Werner in their work \cite{Summers:1987fn,Summers:1987squ}, namely: maximal violation of the Bell-CHSH inequality in wedge regions occurs when the eigenvalue $\lambda\approx 1$ of the modular operator $\delta$ is attained. This is not a mere coincidence. In fact, the eigenvalue $\lambda\approx 1$ has a special meaning: it corresponds to the fixed point of the modular flow $\delta^{it} = e^{it \log(\delta)}$, being deeply connected to the fact that the von Neumann algebra of the local operators in a relativistic Quantum Field Theory is of the type $III_{\lambda=1}$. We recall that the eigenvalues $\lambda^2$ of the modular operator $\delta$ are given by $\lambda^2(\omega) = e^{-2 \pi \omega}$ so that expression \eqref{bellf1} can be rewritten as 
\begin{equation} 
\langle 0| {\cal C} |0\rangle = 2\sqrt{2} \frac{2 \int_0^\infty d\omega \; h(\omega)^2  }{\sqrt{\int_0^\infty d\omega \; h(\omega)^2\left(1 +\lambda(\omega)^2 \right) }{\sqrt{\int_0^\infty d\omega \; h(\omega)^2\left(1 + (\lambda(\omega))^{-2} \right) }} } \;. \label{bellf2}
\end{equation} 
from which the reason for the choice \eqref{abt} becomes manifest. In fact, for small values of the parameters $(a,b)$ expression \eqref{abt} gets concentrated essentially near $\omega \approx 0$, which corresponds precisely to $\lambda^2\approx 1$, where $\bra{0}\mathcal{C}\ket{0} \approx 2\sqrt{2}$. We expect therefore that large violations of the Bell-CHSH inequality will be attained in this region. 

\begin{figure}[h]
\centering
\includegraphics[width=.6\textwidth]{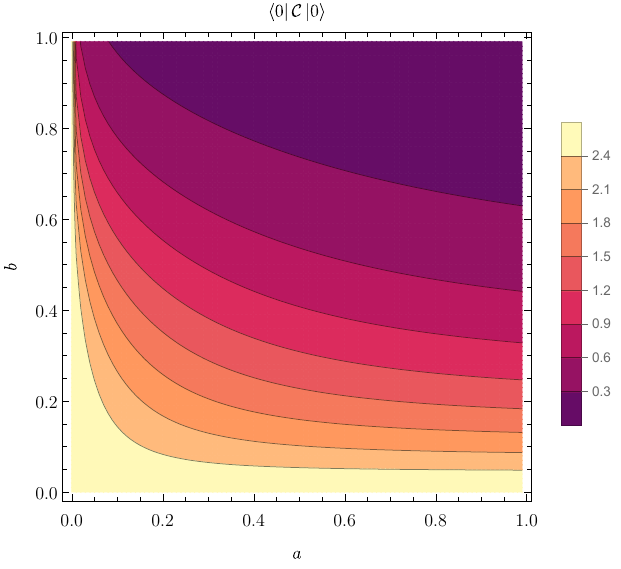}
\caption{Contour plot of the Bell-CHSH correlator from Eq.~\eqref{bellf2} as a function of the parameters $a$ and $b$ of $h(\omega)$ from Eq.~\eqref{phith}. As can be noticed, decreasing either $a$ or $b$ results in an increase in the Bell-CHSH correlator towards the Tsirelson bound of $2\sqrt{2}$.}
\label{fig:fig1}
\end{figure}

\begin{figure}[h]
\centering
\includegraphics[width=\textwidth]{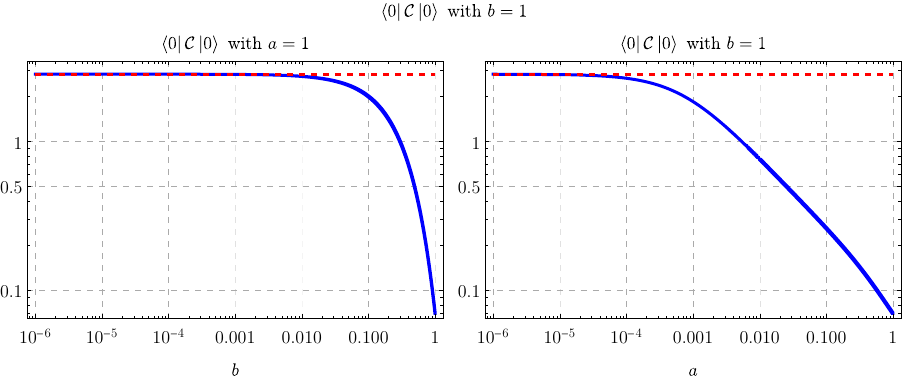}
\caption{The left plot shows the Bell-CHSH correlator from Eq.~\eqref{bellf2} as a function of $b$ for fixed $a = 1$ while the right plot shows the Bell-CHSH correlator as a function of $a$ for fixed $b = 1$. As can be seen, in both plots the Bell-CHSH correlator approaches the Tsirelson bound (red dashed line) of $2\sqrt{2}$ as $a$ and/or $b$ becomes smaller.}
\label{fig:fig2}
\end{figure}

To illustrate this feature, we use the analytical solution to the integral,
\begin{eqnarray} 
\int_0^\infty d\omega \; h(\omega)^2 & = & \frac{\pi a}{2} e^{a^2/b} erfc(a/\sqrt{b}) \;, \label{intgals}
\end{eqnarray}
where $erfc(x)$ denotes the complementary error function 
\begin{eqnarray} 
erf(x) &=& \frac{2}{\sqrt{\pi}}\int_0^x e^{-t^2} dt \;, \nonumber \\
erfc(x) & = & 1-erf(x) =  \frac{2}{\sqrt{\pi}}\int_x^\infty  e^{-t^2} dt \;. \label{erf}
\end{eqnarray}
and numerical solutions to the integrals
\begin{equation}
    \int_0^\infty d\omega \; h(\omega)^2\lambda(\omega)^2 \quad\text{and}\quad \int_0^\infty d\omega \; h(\omega)^2\lambda(\omega)^{-2}.
\end{equation}
These were computed using the Gauss-Kronrod method implemented in Mathematica as \texttt{GaussKronrodRule} with the settings \texttt{AccuracyGoal$\to\infty$}, \texttt{PrecisionGoal$\to8$} and \texttt{WorkingPrecision$\to16$} which internally performs the computations using $16$ digits of precision and produces results with relative errors smaller than $10^{-8}$ in this case. The parameters $a$ and $b$ span the interval $[10^{-6},1]$ in steps of $10^{-2}$ in Fig.~\ref{fig:fig1} and of $10^{-5}$ in Fig.~\ref{fig:fig2}. As can be seen in these figures, the Bell-CHSH correlator $\bra{0}\mathcal{C}\ket{0}$ approaches $2\sqrt{2}$ as either $a$ or $b$ goes to zero.

From the present analysis it follows that many analytic solutions can be found. Essentially, any choice of the function $h(\omega)$ which enables us to stay close to the eigenvalue $\lambda^2\approx 1$ will give rise to violations arbitrarily near to the Tsirelson bound. 
\clearpage

\section{Conclusion }\label{sect6} 

In the present work we have investigated the violation of the Bell-CHSH inequality in the vacuum state of a free massless Majorana field in $1+1$ dimensional Minkowski spacetime. By relying on the introduction of the rapidity variable $\theta$, the 1-particle Hilbert space of the theory has been explicitly identified as $L^2(d \theta, {\mathbb{R}}) \oplus L^2(d \theta, {\mathbb{R}})$. In this construction, the Tomita-Takesaki modular theory was introduced by making use of the Bisognano-Wichmann results for wedge regions. We remark that, to properly handle the fermionic nature of the Majorana field and obtain commuting observables at spacelike separations, the fermion-parity twist was employed. A central result of our analysis is the reduction of the Bell-CHSH correlator to a form which depends entirely on a single real smooth function $h(\omega)$. This feature facilitates the search for simple semi-analytic solutions. As we have seen in Sec.~\ref{sect5}, by judiciously choosing $h(\omega)$ such that its support is concentrated around $\omega \approx 0$, the correlator approaches the Tsirelson bound of $2\sqrt{2}$, corresponding to the maximal violation of the Bell-CHSH inequality.

This near saturation is not a mere coincidence. It explicitly confirms the profound connection, originally pointed out by Summers and Werner, between the maximal violation of the Bell-CHSH inequality and the eigenvalue $\lambda \approx 1$ of the modular operator. As delineated in our discussion, this eigenvalue corresponds to the fixed point of the modular flow and is deeply connected to the fact that the von Neumann algebra of the local operators in a relativistic Quantum Field Theory is of type $III_{\lambda=1}$. From the present analysis, it follows that the framework of modular wedge localization provides a highly tractable and elegant setup to study quantum entanglement in QFT. The present results pave the way for further investigations, such as the extension of this modular approach to other spacetime regions, as well as the study of the effects of interactions and finite temperatures on the saturation of the Tsirelson bound.

\section*{Acknowledgments}
The authors would like to thank the Brazilian agencies CNPq and CAPES for financial support.  S. P.~Sorella is a  CNPq researcher under the contract 302991/2024-7. J. R. A. Quispe acknowledges financial support by Coordenação de Aperfeiçoamento de Pessoal de Nível Superior - Brasil (CAPES) - Finance Code 001.



\begin{thebibliography}{99}



\bibitem{Bell:1964kc}
J.~S.~Bell,
Physics Physique Fizika \textbf{1} (1964), 195-200
doi:10.1103/PhysicsPhysiqueFizika.1.195

\bibitem{Clauser:1969ny}
J.~F.~Clauser, M.~A.~Horne, A.~Shimony and R.~A.~Holt,
Phys. Rev. Lett. \textbf{23} (1969), 880-884
doi:10.1103/PhysRevLett.23.880


\bibitem{Guimaraes:2024mmp}
M.~S.~Guimaraes, I.~Roditi and S.~P.~Sorella,
Rev. Phys. \textbf{13} (2025), 100121
doi:10.1016/j.revip.2025.100121
[arXiv:2410.19101 [quant-ph]].


\bibitem{BESIII:2025vsr}
M.~Ablikim \textit{et al.} [BESIII],
Nature Commun. \textbf{16} (2025), 4948
doi:10.1038/s41467-025-59498-4
[arXiv:2505.14988 [hep-ex]].


\bibitem{Barr:2024djo}
A.~J.~Barr, M.~Fabbrichesi, R.~Floreanini, E.~Gabrielli and L.~Marzola,
Prog. Part. Nucl. Phys. \textbf{139} (2024), 104134
doi:10.1016/j.ppnp.2024.104134
[arXiv:2402.07972 [hep-ph]].

\bibitem{Summers:1987fn}
S.~J.~Summers and R.~Werner,
J. Math. Phys. \textbf{28} (1987), 2440-2447
doi:10.1063/1.527733

\bibitem{Summers:1987squ}
S.~J.~Summers and R.~Werner,
J. Math. Phys. \textbf{28} (1987) no.10, 2448-2456
doi:10.1063/1.527734

\bibitem{Summers:1987ze}
S.~J.~Summers and R.~Werner,
Commun. Math. Phys. \textbf{110} (1987), 247-259
doi:10.1007/BF01207366


\bibitem{Witten:2018zxz}
E.~Witten,
Rev. Mod. Phys. \textbf{90} (2018) no.4, 045003
doi:10.1103/RevModPhys.90.045003
[arXiv:1803.04993 [hep-th]].





\bibitem{Dudal:2023mij}
D.~Dudal, P.~De Fabritiis, M.~S.~Guimaraes, I.~Roditi and S.~P.~Sorella,
Phys. Rev. D \textbf{108} (2023), L081701
doi:10.1103/PhysRevD.108.L081701
[arXiv:2307.04611 [hep-th]].


\bibitem{Dudal:2026eil}
D.~Dudal and K.~Vandermeersch,
[arXiv:2604.05109 [math-ph]].



\bibitem{Caribe:2026mam}
J.~G.~A.~Carib{\'e}, M.~S.~Guimaraes, I.~Roditi and S.~P.~Sorella,
[arXiv:2603.25873 [hep-th]].


\bibitem{Caribe:2026xoa}
J.~G.~A.~Carib{\'e}, M.~S.~Guimaraes, I.~Roditi and S.~P.~Sorella,
[arXiv:2604.18513 [hep-th]].





\bibitem{Bisognano:1975ih}
J.~J.~Bisognano and E.~H.~Wichmann,
J. Math. Phys. \textbf{16} (1975), 985-1007
doi:10.1063/1.522605





\bibitem{Takesaki:1970aki}
M.~Takesaki, {\it Tomita's Theory of Modular Hilbert Algebras and its Applications},
Springer-Verlag, 1970, 
doi:10.1007/bfb0065832

\bibitem{Bratteli:1979tw}
O.~Bratteli and D.~W.~Robinson,
`{\it Operator  Algebras and  Quantum Statistical Mechanics, 1.} , Springer-Verlag (1987)


\bibitem{Summers:2003tf}
S.~J.~Summers,
[arXiv:math-ph/0511034 [math-ph]].

\bibitem{Guido:2008jk}
D.~Guido,
Contemp. Math. \textbf{534} (2011), 97-120
[arXiv:0812.1511 [math.OA]].



\bibitem{Brunetti:2002nt}
R.~Brunetti, D.~Guido and R.~Longo,
Rev. Math. Phys. \textbf{14} (2002), 759-786
doi:10.1142/S0129055X02001387
[arXiv:math-ph/0203021 [math-ph]].


\bibitem{Borchers:2000pv}
H.~J.~Borchers,
J. Math. Phys. \textbf{41} (2000), 3604-3673
doi:10.1063/1.533323

\bibitem{Schroer:2014kya}
B.~Schroer,
SIGMA \textbf{10} (2014), 085
doi:10.3842/SIGMA.2014.085
[arXiv:1407.2124 [math-ph]].


\bibitem{Caribe:2026tfe}
J.~G.~A.~Carib{\'e}, M.~S.~Guimaraes, I.~Roditi and S.~P.~Sorella,
[arXiv:2605.06224 [hep-th]].


\bibitem{Cirelson:1980ry}
B.~S.~Cirelson,
Lett. Math. Phys. \textbf{4} (1980), 93-100
doi:10.1007/BF00417500


\bibitem{Haag:1992hx}
R.~Haag, {\it Local Quantum Physics: Fields, Particles, Algebras}, Springer-Verlag (1992)



\end{thebibliography}
\end{document}